\documentclass{article}

\usepackage{bbding}

\usepackage{makecell}

\usepackage{PRIMEarxiv}

\usepackage[utf8]{inputenc} % allow utf-8 input
\usepackage[T1]{fontenc}    % use 8-bit T1 fonts
\usepackage{hyperref}       % hyperlinks
\usepackage{url}            % simple URL typesetting
\usepackage{booktabs}       % professional-quality tables
\usepackage{amsfonts}       % blackboard math symbols
\usepackage{nicefrac}       % compact symbols for 1/2, etc.
\usepackage{microtype}      % microtypography
\usepackage{lipsum}
\usepackage{fancyhdr}       % header
\usepackage{graphicx}       % graphics
\graphicspath{{media/}}     % organize your images and other figures under media/ folder

\title{Large Language Model-Enhanced Interactive Agent for Public Education on Newborn Auricular Deformities
}

\author{
  Shuyue Wang \\
  Shanghai A\&I Co.Ltd. \\
  \texttt{henri\_w\_91@hotmail.com} \\
  \AND
  Liujie Ren \\
  FPRS Department/ENT Institute, Eye and ENT Hospital, Fudan University \\
  \texttt{renliujie@fudan.edu.cn} \\
  \AND
  Tianyao Zhou \\
  Baidu Inc. \\
  \texttt{ztyernest@163.com} \\
  \AND
  Lili Chen \\
  FPRS Department/ENT Institute, Eye and ENT Hospital, Fudan University \\
  \texttt{lilychen0930@hotmail.com} \\
  \AND
  Tianyu Zhang  \\
  FPRS Department/ENT Institute, Eye and ENT Hospital, Fudan University \\
  \texttt{ty.zhang2006@aliyun.com} \\
  \AND
  Yaoyao Fu \\
  FPRS Department/ENT Institute, Eye and ENT Hospital, Fudan University \\
  \texttt{fuyaoyao2007@126.com} \\  
  \AND 
  Shuo Wang \\
  Digital Medicial Research Center, School of Basic Medical Sciences, Fudan University \\
  \texttt{shuowang@fudan.edu.cn} \\
}

\begin{document}
\maketitle

\begin{abstract}
Auricular deformities are quite common in newborns with potential long-term negative effects of mental  and even hearing problems.
Early diagnosis and subsequent treatment are critical for the illness; yet they are missing most of the time due to lack of  knowledge among parents.
With the help of large language model of Ernie of Baidu Inc., we derive a realization of interactive agent.
Firstly, it is intelligent enough to detect which type of auricular deformity corresponding to uploaded images, which is accomplished by PaddleDetection, with precision rate 75\%.
Secondly, in terms of popularizing the knowledge of auricular deformities, the agent can  give professional suggestions of the illness to parents.
The above two effects are  evaluated via tests on volunteers with control groups in the paper.
The agent can reach parents with newborns as well as their pediatrician remotely via Internet in vast, rural areas with quality medical diagnosis capabilities and professional query-answering functions, which is good news for newborn auricular deformity and other illness that requires early intervention for better treatment.
\end{abstract}

% keywords can be removed

\section{Introduction}

\begin{figure}[htbp]
    \centering
    \includegraphics[width=0.99\textwidth]{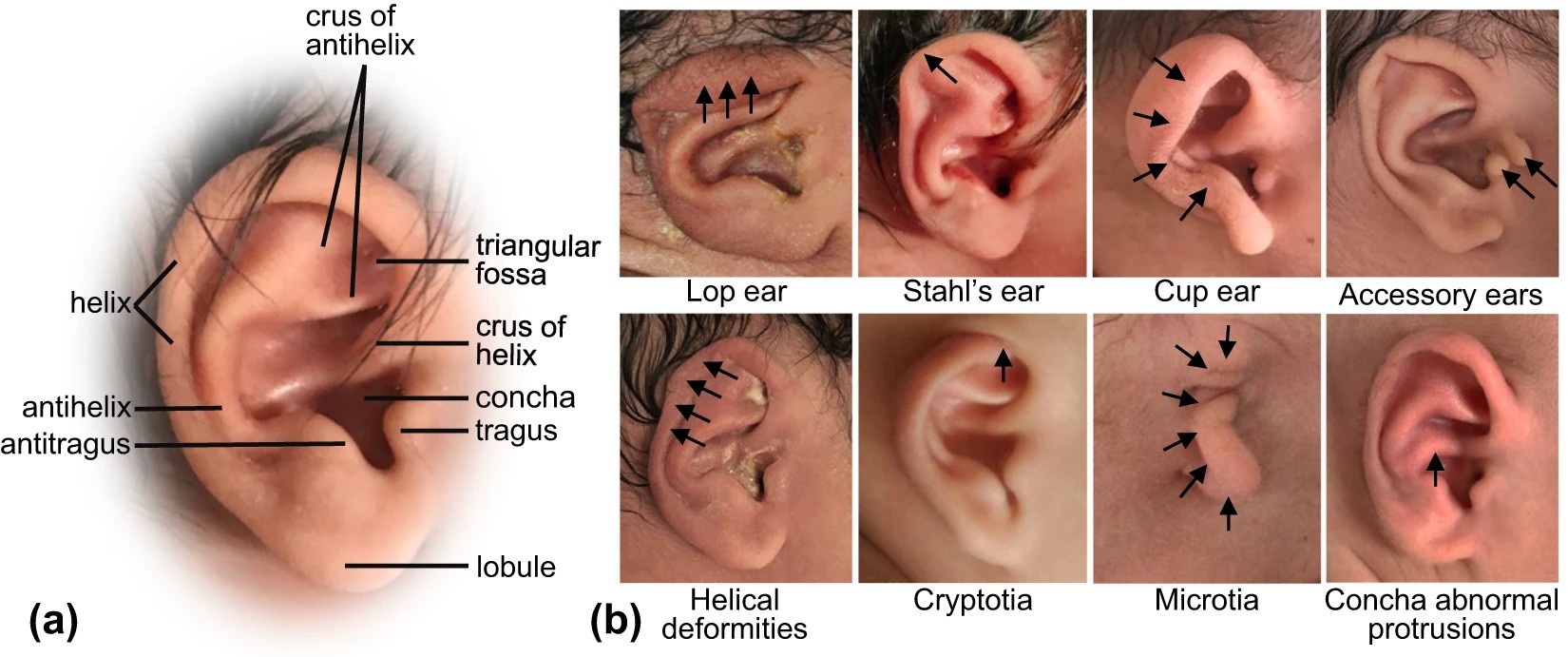}
    \caption{The substructures of the auricle and different types of ear deformities. (a) presents the morphology of a normal ear from one newborn. (b) shows examples of some sub-types of auricular deformities. The abnormal structures are marked with arrows. The illustration is also shown from paper\cite{ren2024publicly}.}
    \label{auricularDeformities}
\end{figure}

% newborn auricular deformities what are they how harmful they are
Auricular deformity ranks among the most prevalent deformities observed in newborns, occurring in approximately 50\% of cases.\cite{zhao2017morphometric}
Its unusual appearance as direct cause of aesthetic concerns may lead to psychological distress, anxiety, feelings of inferiority, and interpersonal difficulties for both children and their parents. 
More severe deformities can be associated with hearing impairments.\cite{jiamei2008investigation}
In fact, only one-third of these anomalies resolve naturally, underscoring the substantial demand for effective corrective methods.\cite{schultz2017newborn}
Figure.\ref{auricularDeformities} illustrates the definition of auricular deformity problem and its main types.

% one important way to prevent its happening   early diagnosis  why is so
Because of its impact not just on the physical appearance but also the psychological development of children,  auricular deformities warrant early intervention rather than being overlooked.\cite{hui2023ear}
Timely and precise diagnosis of ear deformities in newborns is paramount for successful non-surgical corrective treatment.\cite{ren2024publicly}
Conventional approach to treating  auricular deformities involves plastic surgery typically performed after the child reaches the age of 5, by which time the ear has reached approximately 90\% of its adult size. 
Nonetheless, surgical interventions come with drawbacks, including substantial trauma and potential complications, thereby elevating children's discomfort levels and imposing financial burdens on their parents.\cite{hui2023ear}
As far as ear molding is concerned, it is effective, inexpensive, safe, and painless, but it must be administered within a narrow time frame, ideally within 2 to 3 months after birth.\cite{chen2022using}birth.\cite{chen2022using}\cite{nigam2020nonoperative}
It is certain that nonsurgical correction in the early stages offers greater benefits for newborns with congenital auricular deformities.\cite{hui2023ear}

% early interventions
Since addressing auricular deformities consistently necessitates early intervention, there are  two key prerequisites that are involved:
Firstly, widespread awareness among parents of newborns regarding the illness.
From clinical observation, the initiation of treatment for auricular deformities in newborns tends to occur relatively late in China.\cite{PMID:38297872} 
The deficiency in parents' awareness of the condition may stem from widespread misconceptions held by many parents as well as some obstetricians. 
There's a prevalent belief that these deformities will naturally correct themselves with time, leading many children to miss out on valuable opportunities for early nonsurgical correction. 
However, only a minority of newborn auricular deformities heal spontaneously.\cite{hui2023ear}
Secondly, there is a need for an affordable initial diagnostic process; otherwise, the financial and time costs of medical treatment would be prohibitively high. 
For instance, diagnosing auricular deformities often depends heavily on the empirical judgment of clinicians, particularly pediatricians, due to the intricate nature of the human auricle and its multiple sub-structures.

% agent
The paper refers the solution to prevalent auricular deformity issue to Artificial Intelligence in aiding health equity.
One of its essential element is "agent-based system" which is evolving rapidly to address complex or simple issues in tele-health.\cite{clinpract13020042}
These systems provide substantial benefits for pain management and healthcare services by facilitating the physician-patient relationship, particularly in challenging circumstances.
AI can play a pivotal role in leveraging information technology to enhance patient care. 
Various tele-health technologies facilitate the connection between  personnel  and patients to deliver healthcare services, with modalities encompassing synchronous, asynchronous, and remote patient-monitoring methods.\cite{bhatia2022novel}
For example, chatbot-based symptom checker apps have gained popularity in healthcare, engaging users in conversations resembling human interactions and providing potential medical diagnoses.\cite{10.1145/3589959}
The conversational design holds the potential to greatly enhance user perceptions and experiences, consequently improving the medical care they receive. 
Additionally, empirical evidence supports the promotion of consumer-centered AI practices in other contexts.\cite{doi:10.1080/10447318.2022.2095093}
The use of agents enables remote access to academic knowledge and automated diagnosis for illness treatment, thereby inspiring us to apply this technology to address auricular deformities.

% we are now deploying llm to enhance interactive agent for public education on newborn auricular deformities, how is it gonna help enhancement in this thing
We have implemented an interactive intelligent system utilizing the large language model Ernie, developed by Baidu Inc., to improve public awareness regarding newborn auricular deformities and preliminary diagnosis. 
% The project is named "fdear" which is combination of "Fudan University" and "ear".
In this paper, we quantitatively evaluate the system's effectiveness in judging auricular deformity types and qualitatively assess its impact on knowledge dissemination through professional query-response interactions. 
The service of Ernie-fdear is mainly for Chinese patients, therefore the illustrations in the paper are mainly in Chinese, where necessary English translation will be provided.
We anticipate that our proposed approach will enhance the treatment of auricular deformities, with the potential to extend this experience to other areas of newborn illness.

\section{Methods}
\begin{figure}[htbp]
    \centering
    \includegraphics[width=0.99\textwidth]{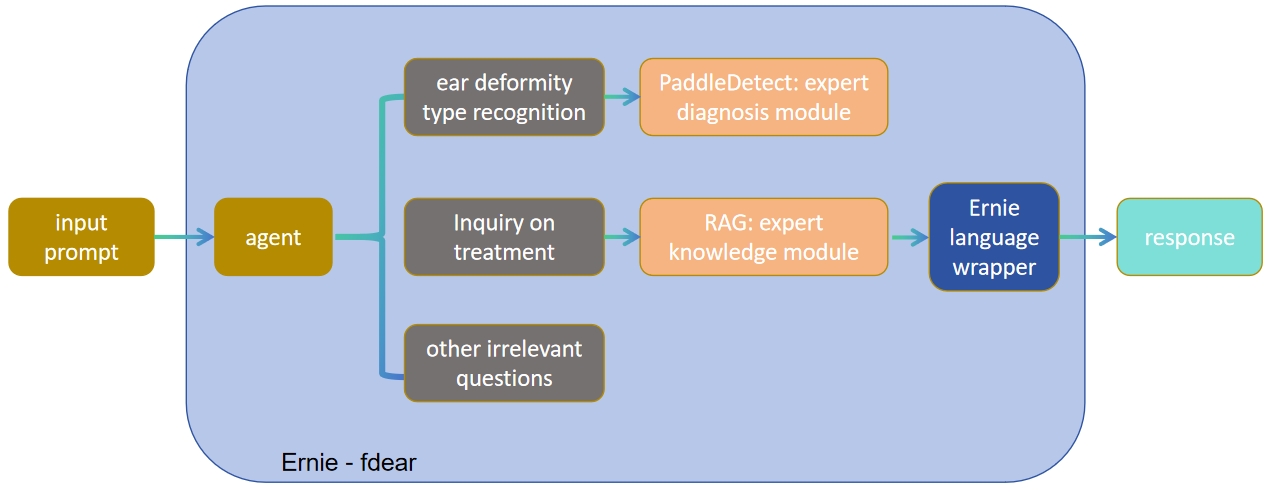}
    \caption{Workflow of the project where the interactive agent accomplishes three functions via three modules.}
    \label{workflo}
\end{figure}

% legal permission and other things ...
% The study was approved by Ethics Committee of School of Basic Medical Sciences, Fudan University.
% The data publication was approved by the committee, and the parents were acknowledged and provided informed consent to the open publication of the anonymized data (including the ear images and the health data). 
In the project, Ernie, the Large Language Model product of Baidu Inc., is deployed with algorithm with software deployment to realize the ‘fdear’system. 
The input of the system is  prompt, and the information is digested  in the system before it gives out response, as is shown in figure.\ref{workflo}.
The prompt is sent to the agent to decide which function should be activated: the main three functions are categorized as follows:

\begin{itemize}
    \item Ear deformity type recognition (cf. table.\ref{tab1}).  If the prompt is an uploaded image, the module tells if there is ear in the picture. If yes, it goes on to classify the ear into normal  or deformed type. This is the expert diagnosis module done via PaddleDetection, a vision detect neural network run by Baidu Inc.
    \item Inquiry on treatment (cf. table.\ref{tab2}). If the prompt is in form of text, the agent will decide if the question is relevant to  auricular deformity. If yes, it goes on to exploit the expert knowledge module  with help of RAG.
    \item Other irrelevant questions. If the prompt is neither relevant auricular image nor inquiries of relevant auricular questions,  the project is detoured letting Ernie does the feedback on its own  capability.
\end{itemize}

A more detailed description of the agent mechanism of Ernie-fdear is in table.\ref{tab1} for image prompts and in table.\ref{tab2} for textual query prompts, where the translation is also adhered in the corresponding places.
Image is the most direct way to interact with our auricular agent.
On one hand, in table.\ref{tab1}, in the first case, a relevant image is input, thus the agent gives the diagnosis of the ear that the ear belongs to the normal type by triggering Expert Diagnosis module that is mentioned in figure.\ref{auricularDeformities}.
In the second case, an irrelevant image of robot is input, thus the agent responded that it cannot give out analysis on auricular deformity and it simply uses the model of Ernie itself.
On the other hand, in table.\ref{tab2}, in the first case, a relevant textual query is input, thus the agent gives relevant information of the question.
The agent answers with a brief introduction of the illness by triggering RAG module in expert diagnosis module (also in figure.\ref{auricularDeformities}).
In the second case, an irrelevant query is input without  any professional information provided in the answer.

\section{Experiments and Results}
As is shown in figure.\ref{workflo}, the main functionality of the Ernie-fdear is focused on expert diagnosis module and expert knowledge module.
The former is activated by the demand in ear deformity type recognition whilst the latter is activated by the demand in answering the inquiry on treatment.
In this section, the  two functions are discussed and evaluated respectively.

% result 1 accuracy of yolov 5  training via paddleDetection
\subsection{Expert diagnosis module: PaddleDetection}
The expert diagnosis module is established upon PaddleDetection.
PaddleDetection is an object detection library based on the PaddlePaddle deep learning framework (https://github.com/PaddlePaddle/PaddleDetection). 
It provides rich implementations of detection algorithms, including Faster R-CNN, Mask R-CNN, YOLOv3, YOLOv4, PP-YOLO, etc., and supports various backbone networks such as ResNet, VGG, MobileNet, etc.
PaddleDetection also provides practical functions such as data augmentation, multi-card training, distributed training, etc., to help users quickly build efficient object detection models.

The dataset is built in this way: for the tag of normal ears and all concerned auricular deformities (i.e. lop ear, Stahl's ear, cup ear, restricted ears, helical deformities, cryptotia, microtia, altogether 8 classes), there are 3852 images in all with left and right ears.
The ear photos were taken within 48 hours after birth, a smartphone (iPhone 6s, Apple) was used. 
Both the left and right ears were shot with a distance of about 30~60 centimeters to ensure a good resolution of the ear and not too much distortions.

The ear (where it be normal of abnormal) occupies approximately 75\% - 90\% of whole photo size which is distributed among 1080x1440, 3456x4608 and 3024x4032, with 3 RGB channels.
A prescribed bounding box is assigned to each image in the dataset, as many studies achieved to recognize the ear location from an image of human head\cite{ganapathi2023survey}
Ignoring thresh is set 0.7.
The base learning rate is 0.001, with epochs set 500.
Validation shows that the category classification accuracy rate hits 75\%; the normal/abnormal recognition accuracy is as high as 90\%.

\subsection{Expert knowledge module: RAG}
Retrieval-Augmented Generation (RAG) is a promising approach for mitigating the hallucination of large language models (LLMs).\cite{jeong2023generative}
It is known as LLM utility in that it is robust in result producing, which combine pre-trained parametric and non-parametric memory for language generation.
In addition, the cost is way cheaper than finetuning of large foundation model, be it global or  partial weight adjusting, because it is based on embedding similarity and prompt template setting.
The plain and scientific document of expert knowledge is quite convenient for the case of RAG.\cite{chen2023benchmarking}
With RAG, a prompt is written  so that its example part contains relevant auricular deformity knowledge surrounding the topic input query by user.
Thus a complete prompt can be input to retrieve professional responses.

In our case, about 75 MegaBytes document is appended to the RAG system developed by Baidu Inc.
In order to judge the effect of the Ernie-fdear query answering, we set up a questionnaire (table\ref{tab3}.
We test 15 volunteers, 5 are professional doctors, 5 are users of Ernie LLM but without permission to use Ernie-fdear, and 5 are users of Ernie-fdear.
They have to answer the questionnaire in their limited reach of information, after reading an article (uploaded in the \href{https://github.com/shuyueW1991/test_article_auricular_deformity}{link})
By giving 1 point for correctly answering 1 question, the average score of the three groups are: 5, 2, 4.5, respectively.

% \section{Examples of citations, figures, tables, references}
% \label{sec:others}
% \lipsum[8] \cite{kour2014real,kour2014fast} and see \cite{hadash2018estimate}.

% The documentation for \verb+natbib+ may be found at
% \begin{center}
%   \url{http://mirrors.ctan.org/macros/latex/contrib/natbib/natnotes.pdf}
% \end{center}
% Of note is the command \verb+\citet+, which produces citations
% appropriate for use in inline text.  For example,
% \begin{verbatim}
%    \citet{hasselmo} investigated\dots
% \end{verbatim}
% produces
% \begin{quote}
%   Hasselmo, et al.\ (1995) investigated\dots
% \end{quote}

% \begin{center}
%   \url{https://www.ctan.org/pkg/booktabs}
% \end{center}

% \subsection{Figures}
% \lipsum[10] 
% See Figure \ref{fig:fig1}. Here is how you add footnotes. \footnote{Sample of the first footnote.}
% \lipsum[11] 

% \begin{figure}
%   \centering
%   \fbox{\rule[-.5cm]{4cm}{4cm} \rule[-.5cm]{4cm}{0cm}}
%   \caption{Sample figure caption.}
%   \label{fig:fig1}
% \end{figure}

% \subsection{Tables}
% \lipsum[12]
% See awesome Table~\ref{tab:table}.

% \begin{table}
%  \caption{Sample table title}
%   \centering
%   \begin{tabular}{lll}
%     \toprule
%     \multicolumn{2}{c}{Part}                   \\
%     \cmidrule(r){1-2}
%     Name     & Description     & Size ($\mu$m) \\
%     \midrule
%     Dendrite & Input terminal  & $\sim$100     \\
%     Axon     & Output terminal & $\sim$10      \\
%     Soma     & Cell body       & up to $10^6$  \\
%     \bottomrule
%   \end{tabular}
%   \label{tab:table}
% \end{table}

% \subsection{Lists}
% \begin{itemize}
% \item Lorem ipsum dolor sit amet
% \item consectetur adipiscing elit. 
% \item Aliquam dignissim blandit est, in dictum tortor gravida eget. In ac rutrum magna.
% \end{itemize}

\section{Conclusion}
In order to reduce the societal impact of auricular deformity among newborns in communities in vast regions of China, an large language model-enhanced interactive agent Ernie-fdear is introduced for public education  as well as preliminary diagnosis.
The main two functions of expert diagnosis module driven by PaddleDetection as well as expert knowledge module driven by RAG are not only described but also evaluated in the paper. 
The education effect also turns out successful since in the test the group using Ernie-fdear performs similarly to expert group. 
The agent can reach  newborns  remotely via Internet  with quality medical diagnosis capabilities and professional query-answering functions, which is good news for newborn auricular deformity  that requires early intervention for better treatment.
We hope this approach can be viewed as part of the efforts to be dedicated to computation equality on the planet.

\section*{Acknowledgments}
Yet another symbolic work representing strong applicability of machine learning upon noble career.

%Bibliography
% \bibliographystyle{unsrt}  
\bibliographystyle{ieeetr}  
\bibliography{references}  

\appendix
\section{Appendix}

\begin{table}
\caption{Input prompts in form of uploaded images with their feedback by Ernie-fdear agent.}\label{tab1}
\begin{tabular}{|l|l|l|}
\hline
Input prompt & Feedback & Explanation\\
\hline
\includegraphics[width=0.15\textwidth]{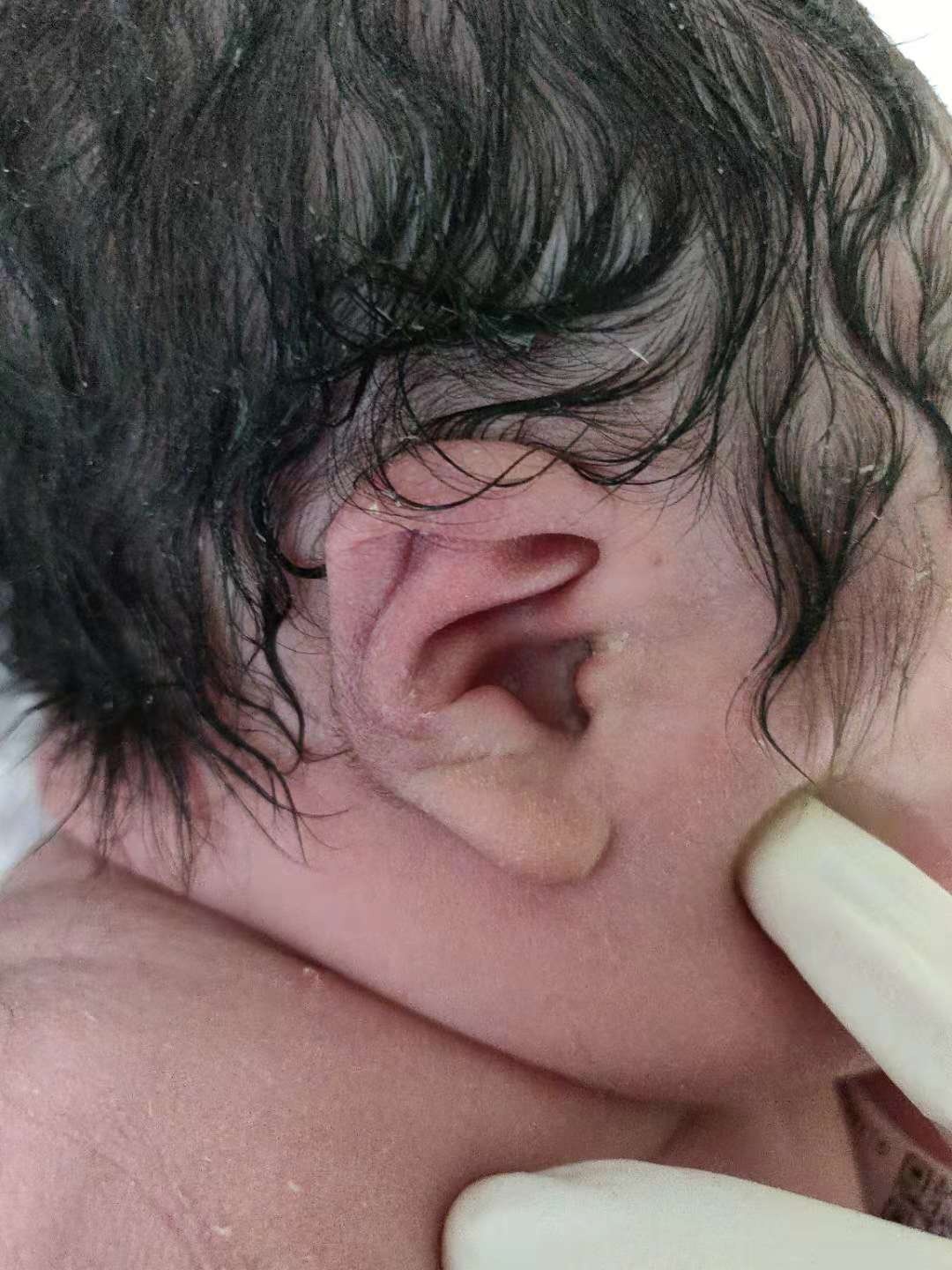}  &  \includegraphics[width=0.6\textwidth]{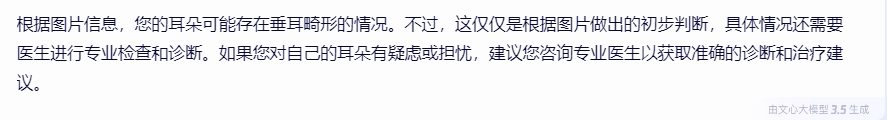}  & \makecell[c]{relevant, \\ Expert \\Diagnosis \\ module \\activated.} \\
   &  \makecell[l]{ (Corresponding translation of the message in the above screenshot:\\ according to the picture, the ear is of 'lopped ear'\\ deformity type.  However, this preliminary  diagnosis \\ requires  professional examination from professional \\ doctors.   You should consult doctors for more  \\details about auricular information.) \\  \\ \\ \\ \\}   & \\ 
\includegraphics[width=0.15\textwidth]{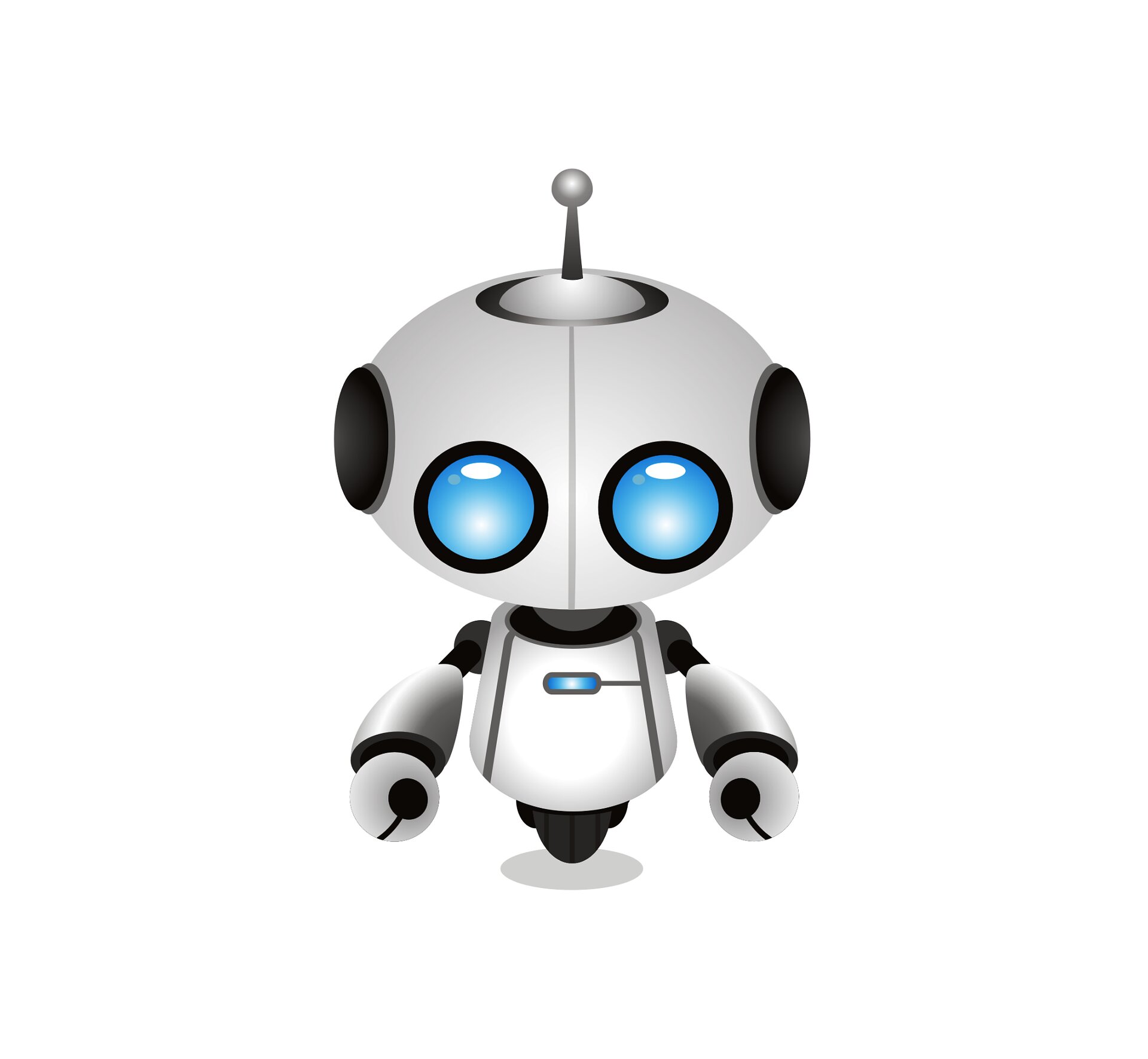}  &  \includegraphics[width=0.6\textwidth]{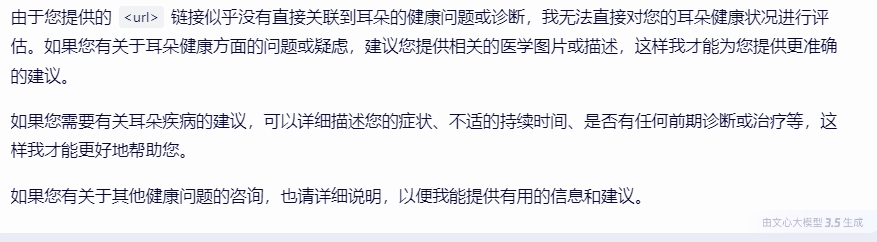} & \makecell[c]{irrelevant,  \\ simply \\ utilize \\ Ernie \\itself.}\\
   &  \makecell[l]{ (Corresponding translation of the message in the above screenshot:\\ your url to the image seems irrelevant to auricular \\ deformity issue. 
   I cannot provide direct evaluation. \\ Relevant medical image of description is suggested \\ for more precise judgment. You should give \\ details about  your symptoms, duration of \\ illness and other relevant information.) \\ } &   \\
\hline
\end{tabular}
\end{table}

\begin{table}
\caption{Input prompts in form of queries with their feedback by Ernie-fdear agent.}\label{tab2}
\begin{tabular}{|l|l|l|}
\hline
Input prompt & Feedback & Explanation\\
\hline
\includegraphics[width=0.15\textwidth]{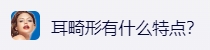}  &  \includegraphics[width=0.6\textwidth]{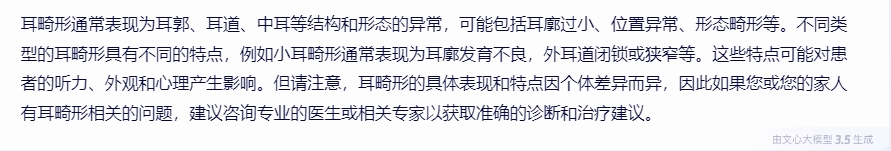} & \makecell[l]{relevant, \\ Expert \\Knowledge \\module \\triggered.}\\
\makecell[l]{ What is\\ auricular \\deformity?}    &  \makecell[l]{ (Corresponding translation of the message in the above screenshot:\\ ear deformities are a type of congenital developmental \\abnormality thatcan result in small ears, abnormal  \\positioning, morphological deformities, and other issues. \\For example, microtia is characterized by poor ear \\development and atresia of the external auditory \\canal, which can affect a patient's hearing and \\appearance. The specific manifestations of\\ these ear deformities vary among individuals, \\so it is recommended to consult a professional\\ doctor if you or your family members \\have concerns about this topic.) \\  \\ \\ \\ \\} &   \\
\includegraphics[width=0.15\textwidth]{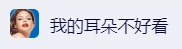}  &  \includegraphics[width=0.6\textwidth]{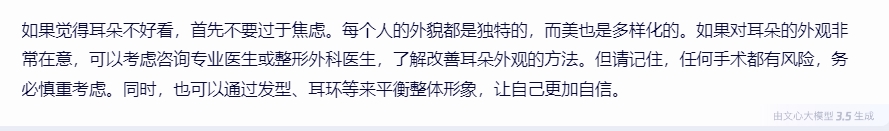} & irrelevant \\
\makecell[l]{ My ear\\ is not  \\pretty.}    &  \makecell[l]{(Corresponding translation of the message in the above screenshot:\\ if you feel that your ears are not aesthetically pleasing, \\it's important not to over-anxietize. Everyone's \\appearance is unique, and beauty is subjective \\and diverse.If you are concerned about the appearance \\of your ears, you can consider consulting a professional \\doctor or plastic surgeon to understand options for \\ improving their appearance. However, please remember\\ that any surgery involves risks, so it's crucial to\\ carefully consider your options. Alternatively, \\you can balance your overall appearance through\\ hairstyles, earrings, and other accessories to enhance\\ your self-confidence.) } &   \\
\hline
\end{tabular}
\end{table}

\begin{table}
\caption{Questionnaire of testing how expert knowledge module trains user to learn auricular deformity issues via RAG.}\label{tab3}
\begin{tabular}{|l|}
\hline
Content of question and choices \\
\hline
    What kind of ears fall into the normal category? \\
    A. Prominent ears\\
    B. Cupped ears\\
    C. Constricted ears\\
    D. Normal substructure ears\\
\\
    Which deformity category is significantly different from the other categories?\\
    A. Microtia\\
    B. Macrotia\\
    C. Lop ear deformity\\
    D. Prominent ear deformity\\
\\
    Which type of deformed ear has poor correction results with ear molds?\\
    A. Lop ear\\
    B. Prominent ear\\
    C. Cryptic ear\\
    D. Microtia\\
\\
    What is the most important determinant for ensuring correction  with ear molds?\\
    A. Home care\\
    B. Early start time\\
    C. Duration\\
    D. Reducing sweating\\
\\
    Which type of ear deformity is prone to accompanying hearing problems?\\
    A. Lop ear deformity\\
    B. Severe prominent ears\\
    C. Microtia\\
    D. Abnormal protrusion of the ear concha\\
\\
\hline
\end{tabular}
\end{table}

\end{document}